\documentclass[onecolumn,
spreprintnumbers,amsmath,amssymb]{revtex4}
\usepackage[dvips]{graphics}
\usepackage{color}

\newcommand{\ee}{\end{eqnarray}}
\newcommand{\be}[1]{\begin{eqnarray} 
\label{#1}
 }

\newcommand{\mtrx}[2]{\left(\begin{array}{#1} #2 \end{array}\right)}
\newcommand{\eref}[1]{(\ref{#1})}
\newcommand{\der}{\mathrm{d}}
\newcommand{\tr}{\mathrm{Tr}\,}

\newcommand\ie {{\it i.e. }}
\newcommand\eg {{\it e.g. }}

\newcommand\grad{\vec\nabla}

\newcommand{\ul}\underline

\newcommand{\GR}{G^\mathrm{R}}
\newcommand{\GoR}{G^{0\mathrm{R}} }
\newcommand{\GA}{G^\mathrm{A}}
\newcommand{\GoA}{G^{0\mathrm{A}} }

\newcommand{\GK}{G^\mathrm{K}}
\newcommand{\GoK}{G^{0\mathrm{K}} }
\newcommand{\omp}{\omega^+}
\newcommand{\prip}[1]{\mathcal{P}\left(#1\right)}

\newcommand{\DT}{\Delta}
\newcommand{\pr}{^\prime}

\renewcommand{\vec}[1]{\text{\boldmath{$ #1 $}}}
\newcommand{\uv}[1]{\vec{\hat{#1}}} 
\newcommand{\ksum}[1]{\int_{#1}}  
\newcommand{\ki}{\ksum{\vec{k\pr}}}
\newcommand{\ti}{\int \frac{\der\theta\pr}{2\pi}}

\newcommand{\Wkk}{W_{\vec{k}\vec{k\pr}}}
\newcommand{\Mkk}{\vec{M}}

\newcommand{\vf}{v_\mathrm{F}}
\newcommand{\erg}{\epsilon}
\newcommand{\aen}{a}
\newcommand{\ben}{b}
\newcommand{\cen}{c}
\newcommand{\sproj}{S}
\newcommand{\idm}{\mathbf{1}}
\newcommand{\ttr}{\tau_\textrm{tr}}
\newcommand{\tilttr}{\tilde{\tau}_{01}}
\newcommand{\taus}{\tau_\textrm{S}}
\newcommand{\tstr}{\gamma}
\newcommand{\tcos}{\tau_\textrm{cos}}
\newcommand{\tsin}{\tau_\textrm{sin}}

\newcommand{\vo}{v_0}
\newcommand{\vecvo}{\vec{v}_0}
\newcommand{\vkm}{\mathbf{v}}

\newcommand{\ordo}[1]{\mathcal{O}(#1)}

\newcommand{\FD}[1]{f_\mathrm{FD}\left(#1\right)}
\newcommand{\FDf}{f_\mathrm{FD}}
\newcommand{\feq}{f^\textrm{eq}}

\newcommand{\DS}{D} 
\newcommand{\grex}{\mathcal{D}}

\newcommand{\JC}{\mathcal{J}}

\newcommand{\Ff}{F}
\newcommand{\Gf}{G}

\newcommand{\nz}{NZ}

\begin{document}

\title{Boltzmann approach to the spin Hall effect revisited and \\
 electric field modified collision integrals}

\author{J. Kailasvuori}
\email {kailas@physik.fu-berlin.de}

\affiliation{Department of Physics,
Freie Universit\"at Berlin\\ Arnimallee 14, 14195 Berlin,
Germany}

\date{\today}

\begin{abstract}
\noindent
The intrinsic contribution to the spin Hall effect in a 2DEG with non-magnetic impurities is studied in a quantum Boltzmann approach. It is shown that if the steady state response is perturbative in the spin-orbit coupling parameter $\lambda$, then the precession term---vital for Dyakonov-Perel relaxation and the key to the spin Hall effect in previous similar  Boltzmann studies---must be left out to first order in spin-orbit coupling. In such a case one would have that to lowest order in the parameters electric field, spin-orbit coupling, impurity strength and impurity concentration there is no intrinsic contribution to the spin Hall effect, not only for a Rashba coupling but for a general spin-orbit coupling. To cover all possible lowest order terms we consider also electric field induced corrections to the collision integral in the Keldysh formalism. However, these corrections turn out to be of second order in $\lambda$. For comparison we derive some familiar results in the case when the response is not assumed to be perturbative in $\lambda$. We also include a detailed discussion of why a relaxation time approximation of the collision integral fails.  Finally we make a comment on pseudospin currents in bilayer graphene.

 \end{abstract}


\maketitle

\section{Introduction}
\noindent
 The idea in  spintronics is to manipulate the electron spin for information storage and transfer, as done with the electron charge in electronics. Possible advantages could for example  be smaller resistive losses, or the additional richness that comes from the  non-scalar nature of spin.
In the study of spin currents particular attention has been given to the spin Hall effect (SHE) in a two-dimensional degenerate electron gas (2DEG) with spin-orbit coupling. Perpendicular to an applied electric field,  opposite spins travel in opposite directions, thus creating a spin current without a net charge current.  Such a spin-current can lead to an accumulation of spin at the edges of a sample, although in contrast to the electrical charge analogy  this is not given.  An individual spin can change its direction, for example due to spin precession, and therefore the accumulated spin polarization is not a conserved quantity.  The SHE opens up one possibility of manipulating spin with electric fields. 

The experiments on the  SHE are few and recent.\cite{Kat, Sih, Wun, Zhao, Val, Kim} 
In contrast, much work have been devoted to the theoretical aspects (see \eg the reviews \onlinecite{Eng, Sch, Sinitsyn}) and even in recent years there has been an intense discussion about the  different mechanisms and how to compare results reached by different formalisms. This discussion is closely related to the one on the anomalous Hall effect (AHE) \cite{KarLut}, for example in the distinction between extrinsic (impurity related, \eg skew scattering,
 side-jump
 etc) and intrinsic (band structure related) mechanisms. The latter could be due to structure inversion asymmetry of the confining potential (leading \eg to a Rashba coupling) or due to bulk inversion asymmetry (allowing for a Dresselhaus coupling).

The earliest theoretical works on the SHE were done by Dyakonov and Perel in 1971 \cite{DyaPer}, who showed that a spin-orbit coupling leads to a spin current perpendicular to the electric field. The name Spin Hall effect was coined in 1999.\cite{Hir} A universal spin Hall conductivity in a 2DEG was proposed.\cite{Zhang, Mur, SinPRL92} However, impuritites were neglected in these studies.  Many studies have later shown that universality is lost when impurities are taken into account.\cite{Mis, Kha, Sug, Shy, CulWin, RaiGor, Ino2003, Ino2004, NomJun, NomSin, RaiSch, Dim, ChaLos} On the other hand, there can still be topological edge modes, responsible for the quantum SHE (see \eg ref.~\onlinecite{Koe}).

The effect of non-magnetic impuritites on the intrinsic contribution to the SHE has been studied using a Boltzmann approach \cite{Mis, Kha, Sug, Shy, CulWin, RaiGor} as well as diagrammatic methods \cite{Ino2003, Ino2004, NomJun, NomSin, RaiSch, Dim, ChaLos} like the Kubo formalism. The diagrammatic methods are more systematic and have a more general range. The Boltzmann approach offers  more intuition, but has typically been implemented by identifying distinct processes or hand-picking different contributions rather than systematically covering all possible contributions through a formal apporach. 
Within the Boltzmann language one finds several different approaches. We are going to use the quantum Boltzmann approach, which treats spin coherently.

This paper is going to deal with the intrinsic spin Hall effect in a steady-state calculation to first order in spin-orbit interaction. We attempt in a Keldysh derivation of the Boltzmann equation to account for all contributions present to lowest order in electric field, in spin-orbit splitting and  in strength and concentration of non-magnetic impurities. (Secs.~\ref{s:sc}-\ref{s:lo} and sec.~\ref{s:sh}.) The spin-orbit interaction $H_\textrm{SO}=\vec{\sigma}\cdot \vec{\ben}$ is chosen to be of the isotropic form  $\vec{\ben}=\ben(k)\uv{\ben}(\theta)$, where the unit vector $\uv{\ben}$ has a winding number $N$, \ie $\hat{\ben}_x+i\hat{\ben}_y= e^{i\theta_0+iN\theta}$ with $\theta_0$ being a constant.

Many recent theoretical studies come to the conclusion that  the spin Hall effect vanishes for $N=\pm 1$ (\eg for the Rashba and linear Dresselhaus spin-orbit couplings) for point-like impurities \cite{Mis, Kha, Sug, Shy, CulWin, RaiGor, Ino2003, Ino2004, NomSin, RaiSch, Dim, ChaLos} as well as finite range impurities \cite{Kha, Sug, Shy, CulWin}. General arguments have been proposed to explain this vanishing.\cite{Dim, ChaLos} (Sec~\ref{s:ga}.) A nonzero result can be found in  refs.~\onlinecite{Ino2004, NomJun}. With an alternative definition of spin currents, a nonzero result is also found in ref.~\onlinecite{Sug}. 

For other odd $N$ the SHE is nonzero \cite{Shy, Sug}, assumes a universal value in a specific limit \cite{Shy}, but depends otherwise on the range of the impurity potential and is therefore in general not universal, though it is independent of the spin-orbit splitting $\ben$, the overall strength of the impurity potential and of the  impurity concentration \cite{Shy}. We reproduce these results  in sec.~\ref{s:pt}. Additionally, we give explicit results on the polarization and spin currents for the components of the spin parallel to the plane and calculate  equilibrium spin currents.

We will also show  that the above results on the SHE are based on the response not being
perturbative in the spin-orbit coupling. If perturbativeness is assumed (sec.~\ref{s:assumption}),  the spin-precession term---seemingly the prerequisite for a spin Hall current---must be left out to linear order in spin-orbit coupling, suggesting that to lowest order the intrinsic spin Hall effect is zero for arbitrary winding $N$ (sec.~\ref{s:bewa}). Possible alternative contributions to the SHE  from electric 
field-induced corrections to the collision integral are discussed in sec.~\ref{s:sh}. A contribution that we believe has not been discussed before in the Boltzmann approach turns out to be the only candidate when the precession term is absent. However, this contribution turns out to be of second order in spin-orbit splitting.   

Leaving out the precession term leads to some formal difficulties in the case $N=\pm 1$. The Boltzmann equation becomes unsolvable. However, this can be remedied by including a small spin relaxation term (sec.~\ref{s:bewa}). This could suggest that for $N=\pm 1$, non-magnetic impurities are not enough for a consistent steady-state solution in the spin polarization in the case that the response is perturbative in the spin-orbit coupling. A comparison of spin densities and spin currents in both phase space and real space is given by table~\ref{t:table}. In this context we also make a comment on pseudospin currents in bilayer graphene. 

The  spin-orbit corrections to the collision integral make the analytical treatment considerably more complicated. For the case $|N|=1$ but not for the case $|N|\neq 1$ this complication is necessary even for a qualitatively correct understanding of the vanishing SHE. In appendix~\ref{a:rta} we discuss the failure of the relaxation time approximation, that in many other problems is useful for a simple qualitative understanding and as a starting point for deriving real space equations for example describing thermoelectric effects. In particular, we contrast with the derivation of the Dyakonov-Perel spin relaxation mechanism.

\section{The model---intrinsic versus extrinsic}\label{s:model}
\noindent
For a semiclassical Boltzmann description  (see \eg  \cite{Zub, Ram, Bro,RamSmi}) one needs the Wigner transformed one-particle Hamiltonian.  For the spin-orbit (SO) coupled electrons we are going to study, it is
\be{ham}
H( \vec{x},\vec{p},t)=\aen(k) +\vec{\sigma}\cdot\vec{\ben}(\vec{k})+e\phi (\vec{x},t) 
\ee
with $e<0$ and $\vec{k}(\vec{x},\vec{p},t)=\vec{p}-e\vec{A}(\vec{x},t)$. We want to describe a 2d system with $\vec{x}$ and $\vec{p}$ chosen to lie in the $x,y$-plane. 
Throughout the paper $\hbar=1$. In absence of spin-orbit coupling the dispersion is given by $\aen \propto k^\zeta$, typically $\aen=k^2/2m$. For the Rashba intercation
 $
\ben:=|\vec{\ben}|=\lambda k$ and  
$\uv{\ben}:=\vec{\ben}/b =\uv{\theta}\, , 
$ 
with $\lambda$ parametrizing the strength of the coupling. However,  we want to consider an arbitrary odd-integer winding number $N$ in  $\vec{\ben}=\ben(k)\uv{\ben}(\theta)$ (with $\hat{\ben}_x+i\hat{\ben}_y=e^{i\theta_0+iN\theta}$). 
The energy bands are $\erg^s_\vec{k}=\aen+s\ben$ with $s=\pm$ giving the sign of the spin along the spin quantization axis $\uv{\ben}$, \ie   $\vec{\sigma}\cdot \uv{b}|\uv{b}s\rangle=s|\uv{b}s\rangle$.

The total Hamiltonian $H_\textrm{tot}=H+H_\textrm{imp}$ also includes an impurity potential $V_\textrm{imp} (\vec{x})=\sum_n U(\vec{x}-\vec{x}_n)$ of charged, non-magnetic impurities at positions $\vec{x}_n$. Including the spin orbit coupling experienced at impurities one has
\be{}
H_\textrm{imp}=V_\textrm{imp} +\lambda_\textrm{ext} \vec{\sigma} \cdot \vec{k}\times \grad V_\textrm{imp}\, .
\ee  
$H$ and $H_\textrm{imp}$ are treated very  differently in the Boltzmann approach. $H$ enters to linear order in the kinetic equation, whereas $H_\textrm{imp}$ is in the Keldysh machinery  turned into an impurity averaged self-energy to  appear to quadratic order in the collision integral. 

A spin-orbit coupling enters both through the {\it intrinsic} (\ie band related) term $\vec{\sigma}\cdot \vec{\ben}$ and in the {\it extrinsic} (\ie impurity related) term $\lambda_\textrm{ext} \vec{\sigma} \cdot \vec{k}\times \grad V_\textrm{imp}$, and the consequences of the two are usually studied separately in the literature. This paper  deals only  with the intrinsic contribution to the spin Hall effect.

\section{Semiclassical description of a spin-orbit coupled system}\label{s:sc}
\noindent
In a Boltzmann description of an electron system with spin, the spatial degrees of freedom are treated semiclassically, whereas the treatment of the spin remains quantum mechanical. The state of the system is given by the $2\times2$-matrix valued distribution function $f_{\sigma\sigma\pr}(\vec{x},\vec{p},t)$, here with the spin index $\sigma=\uparrow_z,\downarrow_z$. It is related to the equal time density matrix $\rho_{\sigma\sigma\pr}(x_1,x_2)|_{t_2 = t_1}=\langle \Psi^\dagger_{\sigma\pr}(\vec{x_2},t_1)\Psi_\sigma (\vec{x_1},t_1)\rangle $ by a   Wigner transformation (see \eg refs.~\onlinecite{Zub, Ram, Bro, RamSmi}).  
In absence of scattering one can derive the Boltzmann equation for $f$   by applying Heisenberg's equation of motion on $\rho(x_1,x_2)$, then identifying $t_2=t_1$, Wigner transforming the result and gradient expanding it to first order.  The approximation to stop at first order in gradient expansion is the  {\it semiclassical} approximation,  which relies on the external perturbations, such as electromagnetic potentials, changing negligibly on length and time scales of  the de Broglie wavelength $\lambda_\mathrm{B}$ and time $ \tau_\mathrm{B}=\lambda_\mathrm{B}/\vf$. 

From the matrix elements of the distribution function $f$  one extracts the densities and current densities of charge and spin. The matrix elements are most conveniently expressed in the   
 decomposition $f=\idm f_0+ \sigma_\mu f_\mu =  f_0 +\vec{\sigma}\cdot\vec{f}$ in Pauli matrices (with $\mu=x,y,z$).  
(Throughout the paper we use the convention of summation over repeated indices.) Furthermore, we find it convenient to decompose the vector $\vec{f}=f_\uv{\ben}\uv{\ben}+f_\uv{\cen}\uv{\cen}+f_z\uv{z}$ in its components along the basis vectors $\uv{\ben} (\theta)$, $\uv{z}$ and $\uv{\cen}(\theta)=\uv{z}\times\uv{\ben}(\theta)$, analogous to the cylindrical basis vectors $\uv{k}(\theta):=\vec{k}/k$, $\uv{z}$ and $\uv{\theta}(\theta):=\uv{z}\times\uv{k}(\theta)$.

 The charge density $e n$ and current density $e\vec{j}$  in phase space are derived from $en=\tr (f \partial H/\partial \phi) $ and $e\vec{j}=-\tr (f\partial H/ \partial \vec{A})$, which yields   
\be{curdef}
\begin{array}{rcccccl}
n(\vec{x},\vec{k},t) &:=&  \tr f & = & 2f_0 & 
= &  n^+ + n^- \\
 j_i(\vec{x},\vec{k},t)&:=&\tr (\vkm_i f) & = & 2 f_0\partial_{k_i}\aen+2\vec{f}\cdot \partial_{k_i}\vec{\ben}
  & = & n^+ v_i^+ + n^-v_i^-+\frac{2N\ben}{k}  f_\uv{\cen}\hat{\theta}_i
\end{array}
\ee
with $i=x,y$. 
Here we introduced the velocity matrices $\vkm_i:=\partial_{k_i}H=\partial_{k_i}\aen +\vec{\sigma}\cdot   \partial_{k_i}\vec{\ben}$.  The spin-independent part of the velocity is $\partial_\vec{k}\aen=:\vecvo$. The band velocities are $\vec{v}^s:=\partial_\vec{k}\erg^s=\langle\uv{b}s|\vkm|\uv{b}s\rangle=v^s \uv{k}$.  The intra-band elements $n^\pm:=\langle\uv{b}\pm|f|\uv{b}\pm\rangle=f_0\pm f_\uv{b}$ give the density of each spin band $s=\pm$. The inter-band elements $\langle\uv{b}\pm|f|\uv{b}\mp\rangle=f_z\pm i f_\uv{\cen}$ are important for the coherent treatment of spin and are, for example, present in the last term of \eref{curdef} \footnote{Note also that this term is equally shared between the two bands;  $\langle\uv{b}s|\frac{1}{2} \left \{\vkm, f\right \} |\uv{b}s\rangle = n^s \vec{v}^s+k^{-1}bf_\uv{c} \uv{\theta} $.    }, containing the Zitterbewegung of the spin-orbit coupled electrons. 

The real space densities are obtained by integrating the phase space densities over momentum, \eg 
\be{}
\vec{j}(\vec{x},t)=\int \frac{\der^2 k}{(2\pi)^2} \vec{j}(\vec{x},\vec{k},t)\, .
\ee
 When not otherwise stated, densities are in this paper always assumed to be phase space densities. 

The spin density, \ie the polarization, is given by $s^\mu=\frac{\hbar}{2}\tr (\sigma_\mu f)=f_\mu$ (with $\hbar=1$).  There is not a unique way to define the spin current because spin polarization is not a conserved quantity. (For a proposal on a conserved spin current, see ref.~\onlinecite{ShiZha}. For its implications on the SHE, see ref.~\onlinecite{Sug}.) When band velocities coincide, \ie $\vec{v}^s=\vecvo$, then it is clearly $\vec{j}^\mu =  f_\mu \vecvo $. For the general case we choose the common definition
\be{spincurr}
j_i^\mu = \frac{1}{4}\tr \sigma_\mu \{ \vkm_i, f\} =f_\mu \partial_{k_i} \aen  +f_0 \partial_{k_i}\ben_\mu
\ee  
(with $\{A,B\}=AB+BA$). 
The spin Hall effect is a real space current of $z$-component spins
\be{she}
\vec{j}^z=\int \frac{\der^2 k }{(2\pi)^2}\, f_z\vecvo=\frac{1}{e}\sigma_\mathrm{SH}\uv{z}\times \vec{E}
\ee 
perpendicular to the an applied electric field $\vec{E}$ along the plane. $\sigma_\mathrm{SH}$ is the spin Hall conductivity.

The Boltzmann equation in matrix form is given by
\be{lhs0}
i[H,f]+\partial_T f+\frac{1}{2}\{{\vkm}_i, \partial_{x_i} f\}+eE_i\partial_{k_i} f -\epsilon_{zij}eB_z \frac{1}{2} \{ {\vkm}_i, \partial_{k_j}f\}= \mathcal{J}[f]      
\ee
where the matrix-valued functional  $\mathcal{J}$ is the collision integral. 
In components it reads (from now on the charge $e$ in $eE$ and $eB$ is absorbed into the fields)
\be{lhs1}
\partial_t f_0 + \partial_{x_i} f_0 \partial_{k_i}\aen +
\partial_{x_i} \vec{f}\cdot  \partial_{k_i}\vec{\ben}+
E_i\partial_{k_i}f_0 +\epsilon_{zij} B_z
 (\partial_{k_i }f_0\partial_{k_j} \aen+\partial_{k_i}\vec{f}\cdot \partial_{k_j}\vec{\ben}) &=&\mathcal{J}_0 
 \nonumber 
\\
2\vec{f}\times \vec{\ben}+\partial_t \vec{f}+ \partial_{x_i} \vec{f} \partial_{k_i}\aen +
\partial_{x_i} f_0  \partial_{k_i}\vec{\ben}+
E_i\partial_{k_i}\vec{f} +\epsilon_{zij} B_z
 (\partial_{k_i }\vec{f}\partial_{k_j} \aen+\partial_{k_i}f_0 \partial_{k_j}\vec{\ben}) &=&\vec{\mathcal{J}}
 \ee
However, by virtue of definition \eref{spincurr} the equations \eref{lhs1} can be compactly written as 
\be{lhs2}
 \partial_t n +\partial_\vec{x}\cdot \vec{j}+
 \partial_\vec{k}\cdot (n\vec{E}+\vec{j}\times\vec{B}) &=& 2\mathcal{J}_0
 \nonumber 
 \\
2(\vec{s}\times \vec{\ben})^\mu +
\partial_t s^\mu + \partial_\vec{x}\cdot \vec{j}^\mu+
\partial_\vec{k}\cdot (s^\mu \vec{E}+\vec{j}^\mu\times \vec{B}) &=& \mathcal{J}_\mu \, .      
\ee
The left-hand side of the first equation is the same as for charged, spinless particles in an electromagnetic field.
 Apart from the spin-precession term,  the second equation is of similar form. This is what one would expect  since the electromagnetic field does not interact with the spin in the considered model but only with the charge that the spin sits on. The spin enters in a non-trivial way only through the precession term and through the collision integral.


\section{Derivation of the collision integral in the Keldysh formalism}\label{s:ci}
\noindent
The presence of, for  example, two-body interactions or disorder averaged impurity interaction is in the Boltzmann approach described by the collision integral $\mathcal{J}$. It is assumed that one is in the {\it kinetic regime}, where the de Broglie wavelength $\lambda_\mathrm{B}=1/k_\mathrm{F}$ is much shorter than the scattering length $\ell$.
 Like ref. \onlinecite{Shy} we use the Keldysh formalism (see \eg refs.~\onlinecite{Zub, Ram, Bro, RamSmi}), but among other general methods we can mention the Nonequilibrium statistical operator formalism \cite{Zub}. 
For disorder averaged impurities, see also the compact derivation in ref.  \onlinecite{CulWin}.   

The Keldysh derivation of the semiclassical equations starts with relating $f$ to the Wigner transformed Keldysh Green's function
\be{}
f(\vec{x},\vec{p},t) = \frac{1}{2}+ \int\frac{ \der \Omega}{4\pi i}\,  \GK(\vec{x},\vec{p},t,\Omega)\, .
\ee
The equation of motion for $\GK$ is given by the Dyson equation. Integrating the equation over the frequency $\Omega$  results in the semiclassical Boltzmann equation with the collision term.  This should be contrasted with the {\it quasiclassical} Boltzmann approach (see \eg ref.~\onlinecite{RamSmi}) where the integration is instead performed over $|\vec{k}|$ to obtain a distribution function  $f(\vec{x}, \uv{p},t,\Omega)$. This is the approach for example in refs.~\onlinecite{Shy} and \onlinecite{RaiGor}. 

The Keldysh equation derived from the Dyson equation can be written in two equivalent forms
\be{dyson}
 \hat{\idm}=\left(i\partial_t-H-\Sigma\right)*G \hspace{1cm}\textrm{or}
\hspace{1cm}
  \hat{\idm}=G*\left(i\partial_t-H-\Sigma\right) \, .
\ee
The product involved here is the convolution product, the identity stands for $\hat{\idm}:= \idm \delta\left(\vec{x}_1-\vec{x}_2\right)\delta\left(t_1-t_2\right)$ and quantities are written in Keldysh matrix space with 
\be{}
G=
\begin{pmatrix} \GR & \GK \\0& \GA \end{pmatrix} \hspace{1cm}\textrm{and}\hspace{1cm}
i\partial_t-H-\Sigma = \begin{pmatrix}i\partial_t-H-\Sigma^\mathrm{R}&-\Sigma^\mathrm{K}\\0&i\partial_t-H-\Sigma^\mathrm{A}\end{pmatrix}\, .
\ee
Each element in these matrices is an infinite-dimensional matrix  in real space indices, and in our case also a $2\times2$ matrix in spin indices. 

The two equations in \eref{dyson} contain the same information. To derive kinetic equations one takes the difference of them, which for the Keldysh component yields 
\be{gkeq}
(i\partial_t -H)*\GK-\GK* (i\partial_t-H) = \Sigma^\mathrm{R}* \GK -\GK* \Sigma^\mathrm{A}-\GR *\Sigma^\mathrm{K}+\Sigma^\mathrm{K}*\GA\, .
\ee
The right hand side is going to give the collision integral, which we only treat to lowest order in impurity concentration and impurity strength (first Born approximation),
\be{}
\Sigma_\vec{p}=  
n_\textrm{imp} \int \frac{\der^2 p\pr}{(2\pi)^2} |U(|\vec{p}-\vec{p\pr}|)|^2 G_\vec{p\pr}^0\, ,
\ee
 with $G^0=G(\Sigma=0)$.  ($\Sigma$ is diagonal in momentum space due to the disorder averaging of the impurity interaction.) According to self-consistent Born-approximation we replace $\GoK$ by $\GK$. 
 A crucial approximation comes with choosing the generalized Kadanoff-Baym Ansatz \cite{LipSpiVel} 
\be{gkba}
\GK =  i ( \GR*h-h*\GA)+\ldots
\ee
 where $h:=\GK|_{t_2=t_1}$, which generalizes the quasiparticle approximation $\GK=h(\vec{x},\vec{p},t)\delta(\Omega-\erg)$ for spinless electrons. The approximation can be considered to be an expansion in relaxation times of the system and corresponds to a Markov approximation.

The convolution product takes after Wigner transformation
$ A(\vec{x}_1,t_1,\vec{x}_2,t_2)\rightarrow A(\vec{x},\vec{p},t,\Omega)$ the form $
A*B=A e^{\frac{i}{2}\grex}B$ 
with the Poisson-bracket-like gradient $\grex$.
In a gauge invariant treatment valid when the electromagnetic fields are weak and vary slowly (see \eg p. 344, Vol. 1 in ref. \onlinecite{Zub}), one introduces $\vec{k}(\vec{p},\vec{x},t)=\vec{p}-\vec{A}$ and $\omega (\Omega,\vec{x},t)=\Omega-\phi$ and lets $\{\vec{x},\vec{k},t,\omega\}$ become the new set of independent variables (\ie $\partial_{x_i}\vec{k}=0$). This  changes the gradient into
\be{gigrad}
\grex=
\overleftarrow{\partial}_{x_i}\overrightarrow{\partial}_{k_i} -
\overleftarrow{\partial}_{k_i}\overrightarrow{\partial}_{x_i} +
\overleftarrow{\partial}_\omega\overrightarrow{\partial}_t-
\overleftarrow{\partial}_t\overrightarrow{\partial}_\omega+
E_i (\overleftarrow{\partial}_\omega\overrightarrow{\partial}_{k_i} -
\overleftarrow{\partial}_{k_i}\overrightarrow{\partial}_\omega)+
\epsilon_{ijl}B_i \overleftarrow{\partial}_{k_j}\overrightarrow{\partial}_{k_l}\, 
\ee
with $X\overleftarrow{\partial}Y := (\partial X)Y$ and $X\overrightarrow{\partial}Y := X(\partial Y)$.  Gradient expanding the left hand side of \eref{gkeq} to first order and integrating over the frequency yields the left-hand side of \eref{lhs0}. The right hand side, the collision part,  is usually taken to zeroth order in gradient expansion.
 Inserting \eref{gkba} into \eref{gkeq}, and using that combinations such as $\int \der \omega \, \GR [\ldots] \GR$ vanish, one arrives at the collsion integral (with  $\Wkk:=2\pi n_\textrm{imp}|U(|\vec{k}-\vec{k\pr}|)|^2$)
\be{jc}
\JC & = &-\ki
\frac{\Wkk 
}{2\pi}\int \frac{\der \omega}{2\pi}
(\GR_\vec{k} \DT f  \GA_\vec{k\pr} +\GR_\vec{k\pr} \DT f \GA_\vec{k} )=
\nonumber 
\\
& = &  
-\ki
\frac{\Wkk 
}{2\pi}\int \frac{\der \omega}{2\pi}
(\GoR_\vec{k} \DT f  \GoA_\vec{k\pr} +\GoR_\vec{k\pr} \DT f \GoA_\vec{k} ) +\ldots
\ee
where in the last row only terms of second order in the interaction strength were kept. The shorthand notations
$\DT f=f(\vec{k},\vec{x},t)-f(\vec{k\pr},\vec{x},t)$ and  $\int \frac{\der^2 k\pr }{(2\pi)^2} =:\ki$ were introduced.

In the expression \eref{jc} we also need the retarded and advanced components.  For this one should take the {\it sum} of the two equations \eref{dyson}, which for $\GR(\Sigma=0)$ after Wigner transformation leads to
\be{greq}
2=(\omp -H^0)e^{\frac{i}{2}\grex}\GoR+\GoR e^{\frac{i}{2}\grex}(\omp-H^0) \, ,
\ee
with  $\omp:=\omega +i\eta$ (to take care of the boundary conditions provided by the imaginary part of $\Sigma^\mathrm{R}$ when $\Sigma\neq 0$) and with $H^0:=H-\phi$. To zeroth order in gradient expansion it is solved by 
\be{gor}
\GoR =\sum_{s=\pm} \frac{\sproj_{\uv{\ben}s }}{\omp-\erg^s} 
\hspace{2cm}
\sproj_{\uv{\ben}s}:= 
\frac{1}{2}(\idm+\vec{\sigma}\cdot s\vec{\ben})
\ee  
where $\sproj$ is the spin projection operator.  With this zeroth order $\GoR$, the last line in \eref{jc} can be reformulated to be identical to the result derived in ref. \onlinecite{CulWin} and essentially also to the one derived in ref. \onlinecite{Shy}. 

To the knowledge of the author,  one finds  in the literature always the zeroth order solution for $\GoR$, maybe because in the common cases like spinless electrons or Zeeman-coupled electrons (with a constant magnetic field) the first order contribution to \eref{greq} (and thus to $\GoR$) vanishes. However, for a spin-orbit coupled system it does not. We find the first order contribution 
\be{gor2}
\delta \GoR & =& \sigma_z Nb \frac{E_\uv{\theta} \ben -B_z (\omega-\aen)\partial_k \ben}{k(\omp-\erg^+)^2(\omp-\erg^-)^2}
= [B_z=0]\nonumber \\
& = &\sigma_z N\frac{E_\uv{\theta} } 
{4k}\sum_s (-s\ben^{-1} -\partial_\omega ) \frac{1}{\omp-\erg^s} \, ,
\ee 
with $E_\uv{\theta}:=\vec{E}\cdot \uv{\theta}$ in  the polar decomposition $\vec{E}=E_\uv{k}\uv{k}+E_\uv{\theta}\uv{\theta}$.
In section~\ref{s:sh} we show how the contribution \eref{gor2} modifies the collision integral.  
We will also investigate, whether the corresponding correction could be an alternative source to the SHE not relying on the precession term.\footnote{Note that if one in contrast to equation \eref{greq} takes the {\it difference} of the equations \eref{dyson}, the equation for $\GoR$ is identical to \eref{gkeq} with $\Sigma=0$. This equation does not determine $\GoR$, but given \eref{greq}, the precession term (in the equation for $\GoR$)  implies the presence of \eref{gor2} to first order in electromagnetic fields. }

\section{Collision integral to linear order in spin-orbit coupling}
\label{s:lo}
\noindent
In this section the Boltzmann equation is expanded to first order in the spin-orbit coupling,  as done in refs. \onlinecite{Shy} and \onlinecite{CulWin}, hence assuming $\ben(k_\mathrm{F})\ll \erg_\mathrm{F}$. The subscript F indicates the value of the corresponding quantity at the Fermi surface determined by $\aen=\erg_\mathrm{F}$, where $\erg_\mathrm{F}=\mu$ at low temperatures $k_\mathrm{B}T\ll \erg_\mathrm{F}$.   

The collision integral is taken to the habitual zeroth order in gradient expansion, meaning \eref{jc} with \eref{gor}. To slim down the often lengthy expression for collision integrals, some more shorthand notation is introduced.  $x\pr$ means that the quantity $x$ depends on primed variables such as $\vec{k\pr}$, $s\pr$ etc, whereas $x$ correspondingly depends on $\vec{k}$, $s$. For example $\sproj\pr = \frac{1}{2}(\idm+\vec{\sigma}\cdot s\pr\uv{\ben}_\vec{k\pr})$. Also, $\DT x := x-x\pr$, for example $\DT \erg= \erg_\vec{k}^s-\erg_\vec{k\pr}^{s\pr}$ or $\DT (sb)=sb-s\pr b\pr$.  

Inserting \eref{gor} into \eref{jc} gives 
\be{jcb}
\mathcal{J}_0
&=& -\ksum{\vec{k\pr}} \Wkk\,\, 
\frac{1}{2}\sum_{s s\pr} \delta (\DT\erg )
\left[ 
\frac{1+s s\pr \uv{\ben}\cdot\uv{\ben \pr}  } {2} 
\DT f_0 + 
\frac{s  \uv{\ben}+s\pr \uv{\ben\pr}  } {2} 
\cdot \DT \vec{f}
  \right]+X_0
    \nonumber
    \\
\vec{\mathcal{J}}
&=&- \ksum{\vec{k\pr}} W_{\vec{k}\vec{k\pr}  }\,\, 
\frac{1}{2}\sum_{s s\pr} \delta (\DT\erg )
\left[ 
\frac{1+s s\pr  \mathrm{B}_{\vec{k}\vec{k\pr} }   } {2} 
\DT \vec{f}  +
\frac{s  \uv{\ben}+s\pr \uv{\ben\pr}  } {2} 
\DT f_0
  \right] +\vec{X}\, .
\ee
with the matrix $
\mathrm{B}_{\vec{k}\vec{k\pr}}: = 
\uv{\ben} ( \uv{\ben\pr})^\mathrm{T}+
\uv{\ben\pr} ( \uv{\ben})^\mathrm{T}-
\uv{\ben}\cdot  \uv{\ben\pr}
$.\footnote{
The structure of $\mathrm{B}$ is relevant for graphene (one Dirac cone). Not surprisingly, an equivalent strucutre is found in ref.~\onlinecite{CulWin2} 
We note that one gets a different matrix $\mathrm{B}$ if one instead uses the  nonequilibrium statistical operator method, used for example in ref.~\onlinecite{AusKat}. 
 This difference, to be discussed elsewhere (J. Kailasvuori and M. L\"uffe, {\it in preparation}), has no influence on the present paper since the term involving $\mathrm{B}$ does not contribute to first order in spin-orbit coupling.    
}
 Not written out in equation \eref{jcb} are the principal part terms
\be{}
X=\ki \Wkk \frac{1}{2\pi}\sum_{ss\pr} \prip {\frac{1}{\DT \erg }} \left[
\frac{ss\pr \uv{b}\times\uv{b\pr}}{2}\cdot 
\left(\DT \vec{f} -\vec{\sigma}\DT f_0 \right)-\vec{\sigma}\cdot\frac{\DT (s\uv{b})}{2} \times \DT \vec{f} \right]
\ee
which are not considered to be a part of the elastic collision integral but to be related to renormalization corrections. Such terms can usually left out if one is  interested in the interaction only to lowest order (however, see ref.~\onlinecite{AusKat}) and it is beyond the scope of the present paper to discuss them.  See ref.~\onlinecite{LipMorSpi} for a lucid treatment on  how to interpret and handle them in the case of spinless electrons interacting through a two-body interaction. See also the derivation of Bloch spin relaxation equations in ref.~\onlinecite{LagWil}.  



The delta functions $\delta (\erg_\vec{k}^s -\erg_\vec{k\pr}^{s\pr})$ connecting Fermi surfaces at different $|\vec{k}|$ make it difficult to find an analytical solution. However, in the considered limit $\ben (k_\mathrm{F})\ll \erg_\mathrm{F}$ one can use the expansion
\be{deltexp}
\delta (\DT \erg ) =\delta (\DT \aen) +
\DT(s\ben) \, \delta\pr (\DT \aen ) +\ordo{\lambda^2}
\ee
after which one is left with the spin-independent delta function $\delta (\aen_k-\aen_{k\pr})$ implying $|\vec{k\pr}|=|\vec{k}|$. With this expansion the collision integral reads 
\be{jlin}
\mathcal{J}_0
&=&
 -\ksum{\vec{k\pr}} \Wkk\,\, [
 \delta (\DT\aen )
\DT f_0 + \overbrace{
\delta\pr (\DT \aen)  \DT \vec{\ben} \cdot \DT \vec{f}}^{\ordo {\lambda^2}}
  ]
    \nonumber
    \\
\vec{\mathcal{J}}
&=& -\ksum{\vec{k\pr}} \Wkk\,\, \left[
 \delta (\DT\aen )
\DT \vec{f} + 
\delta\pr (\DT \aen)  \DT \vec{\ben} \DT f_0
  \right]\, .
\ee
The terms with $\delta\pr (\DT \aen)=-\partial_{\aen\pr} \delta (\DT \aen)$ are made sense of by integration by parts.

 If $\vec{f}=\ordo {\lambda}$ (\ie if the polarization vanishes as $\lambda\rightarrow 0$), the term indicated as of order $\ordo{\lambda^2}$ can be neglected to linear order in  $\lambda$. This resulting collision integral is essentially the one found in ref. \onlinecite{CulWin} and is similar to the one in ref.~\onlinecite{Shy}.\footnote{The quasiclassical approach of ref.~\onlinecite{Shy} makes a straightforward comparison not obvious,  but it  seems like there is no analogue  of the term containing $\partial_{\aen\pr}\DT f_0$ that we obtain  after integration by parts.} 
For a physical interpretation of spin-orbit coupling dependent contributions, see ref.~\onlinecite{Shy}.

In section \ref{s:sh} we discuss corrections to the collision integral  when one goes beyond zeroth order in gradient expansion.

\section{Solving the Boltzmann equation}\label{s:pt}
\noindent
We now set out to solve the uniform, steady state Boltzmann equation 
\be{eboltzn}
2\vec{\sigma}\cdot \vec{f}\times\vec{\ben} +\vec{E}\cdot \partial_\vec{k} f  
&=& -\ksum{\vec{k\pr}} \Wkk\,\, \left[
 \delta (\DT\aen )
\DT f + 
\delta\pr (\DT \aen)  \DT \vec{\ben}\cdot \vec{\sigma} \DT f_0
  \right]
\ee
without implementing the  assumption of perturbativeness in $\lambda$, to be discussed in sec~\ref{s:assumption}.  The distribution function $f=f^\textrm{eq}+f^{(E)}$ is linearized only in the electric field and not in the spin-orbit coupling. 

The
distribution $n^\pm = f_0\pm f_\uv{b}$ of respective spin band is in equilibrium given by the Fermi-Dirac (FD) distribution. The equilibrium distribution is therefore given by
\be{equil}
  f^\textrm{eq}_0\pm f^\textrm{eq}_{\uv{\ben}}  = \FD{\erg^\pm-\mu}\hspace{1cm}    \textrm{i.e.}\hspace{1cm}  f^\textrm{eq}=\sum_{\pm}\sproj_{\uv{\ben}\pm}\FD{\erg^\pm-\mu}\, ,    
\ee
with vanishing inter-band elements $f^\textrm{eq}_{\uv{\cen}}=f^\textrm{eq}_z=0$.  Time reversal symmetry requires the  {\it real space} equilibrium polarization  $\vec{f}^{\textrm{eq}}=\int_\vec{k} \vec{f}^{\textrm{eq}}(\vec{k})$ to be  zero, which also follows trivially from the vanishing angular part of the integral. The real space spin current, on the other hand,  need not vanish since it is even under time reversal symmetry. For $|N|\neq  1 $ the real space spin current is trivially zero, but for $N=\pm 1 $ we find  $j_y^x=\mp j_x^y=-m^2\lambda^3/2\pi  +\ordo{\lambda^5} $ for a quadratic dispersion $\aen=k^2/2m$.  

The equilibrium distribution is from now on taken to linear order in $\lambda$, \ie
\be{equil2}
\begin{array}{rcl}
f_0^{\textrm{eq}}&=& \sum_\pm \frac{\FD{\erg^\pm-\mu} }{2}=\FD{\aen-\mu} + \ordo{\lambda^2}
\\
f_\uv{\ben}^{\textrm{eq}}&=&\sum_\pm \pm \frac{\FD{\erg^\pm-\mu} }{2} =\ben \partial_\aen \FD{\aen-\mu} +\ordo{\lambda^2}
\end{array}
\Longrightarrow 
\begin{array}{rcl}
f_0^{\textrm{eq}} &= & \FDf
\\
\vec{f}^{\textrm{eq}} & = & \vec{\ben}\partial_\aen \FDf
\end{array}
\ee
where  $\FDf\equiv  \FD{\aen-\mu}$ from now on. From \eref{equil2} one sees that a small spin-orbit coupling does not change the charge density but induces a small polarization at the Fermi surface. The distribution $f^\textrm{eq}=\FDf+\vec{\sigma}\cdot\vec{\ben}\partial_\aen \FDf
$   satisfies \eref{eboltzn} for $\vec{E}=0$. 

The charge part of \eref{eboltzn} does not depend on the polarization. When a uniform, static electric field is applied, one finds the usual solution  $f_0^{(E)}=-\ttr E_\uv{k}\partial_k \FDf$ where $\ttr$ is the transport relaxation time. Gathering the known terms on the left-hand side one can  write the polarization part of the equation as 
\be{inc1}
\vec{E}\cdot\partial_\vec{k} \vec{f}^\textrm{eq} + \ki W_{\vec{k}\vec{k\pr}}
\delta\pr (\DT \aen) 
 \DT \vec{\ben}  
\DT f_0^{(E)}
= 2\vec{\ben}\times\vec{f}^{(E)} 
  -\ki W_{\vec{k}\vec{k\pr}}
\delta (\DT \aen) \DT \vec{f}^{(E)}\, .
\ee
It turns out (see the appendix) that the equation can be written in the form
\be{fLE}
\Ff
(E_\uv{k}\uv{\ben}-E_\uv{\theta} \uv{\cen} ) +
\Gf
(E_\uv{k}\uv{\ben}+E_\uv{\theta} \uv{\cen}  )
 =2\vec{\ben}\times\vec{f}^{(E)} 
 -
 \int_{\theta\pr} K  \DT \vec{f}^{(E)}|_{k\pr=k}
\, ,
\ee
with the shorthand $\int_{\theta\pr}:= \ti$. 
The functions $\Ff$ and $\Gf$ depend on $k$ and are proportional to $\lambda$. They do not depend on $\theta$, on $\vec{E}$,  on the impurity concentration  $n_\textrm{imp}$ or on the overall strength of the impurity potential $U$. However, they depend on the range of the potential through dimensionless fractions of the Fourier components of the functions  
\be{}
K(k,\DT \theta) &:=&\DS (\aen) \Wkk |_{k\pr=k}=:\sum_m e^{im\DT\theta} K_m
 \nonumber\\
\tilde{K}(k,\DT \theta)&:=&\DS (\aen) kv_0 [\partial_{\aen\pr}\Wkk ]_{k\pr=k} =:\sum_m e^{im\DT\theta} \tilde{K}_m
\, .\ee\footnote{To compare with ref. \onlinecite{Shy}, note that $\partial_{\aen\pr}\Wkk|_{k\pr=k} =  (v_0k)^{-1}\tan^2 (\DT \theta/2) \partial_{\theta\pr} \Wkk|_{k\pr=k}$ .} Particularly, for $N=1$ we find  for a general dispersion $\aen \propto k^\zeta$ that
\be{FandG}
\Ff &=&\lambda \left [
\frac{ \partial_\aen \FDf}{2} 
\left( 
(\zeta-2)\ttr^{-1}
+ \zeta\tau_{12}^{-1} 
-2\ttr \tau_{12}^{-1}{\tilttr}^{-1}
-\tilttr^{-1}+\tilde{\tau}_{12}^{-1}
\right)\ttr +
 \frac{k \partial_k \partial_\aen \FDf}{2} \left( 1+\ttr \tau_{12}^{-1}    \right) \right]
\nonumber \\
\Gf  & =& 
0
\ee
 with $\ttr^{-1}=K_0-K_1$ and introducing shorthands 
$\tau_{12}^{-1}:=K_1-K_2$, $\tilttr^{-1}:=\tilde{K}_0-\tilde{K}_1$ and 
$\tilde{\tau}_{12}^{-1} =\tilde{K}_1-\tilde{K}_2$. We find that $\Gf=0$ also when $\ben$ is not proportional to $k$. The vanishing of $\Gf$ is going to imply the vanishing of the spin Hall effect for the Rashba coupling. 

The combination $E_\uv{k}\uv{\ben}\mp E_\uv{\theta} \uv{\cen}$ has a winding $N\pm 1$. Particularly, for the Rashba case $\uv{\ben}=\uv{\theta}$ (\ie $N=1$)
one has 
\be{angindep}
E_\uv{k}\uv{\ben}-E_\uv{\theta} \uv{\cen} =
\left(\uv{\ben} (\uv{k})^\mathrm{T}-\uv{\cen} (\uv{\theta})^\mathrm{T}\right)\vec{E}=
\mtrx{cc}{
-\sin 2\DT \theta  &  \cos 2\DT \theta \\
\cos 2 \DT \theta & \sin 2 \DT \theta \\}\vec{E} 
\hspace{1cm}
E_\uv{k}\uv{\ben}+E_\uv{\theta} \uv{\cen} =
\mtrx{cc}{
0  & -1 \\
1 & 0}\vec{E} \, ,
\ee
\ie the left-hand side of \eref{fLE} has an angularly independent term. 

The solution can be found by Fourier decomposition in the basis $\{ \uv{\ben},\uv{\cen},\uv{z} \}$
\be{decomp}
\vec{f}^{(E)}=\sum_n e^{in\theta} 
(\uv{\ben} f_{\uv{\ben}n}+  
\uv{\cen} f_{\uv{\cen}n}+
\uv{z} f_{zn}
) =\sum_n e^{in\theta} \mtrx{ccc}{\uv{\ben} & \uv{\cen} & \uv{z}} 
\mtrx{c}{ f_{\uv{\ben}n} \\ f_{\uv{\cen}n} \\ f_{zn} } 
\ee
where the Fourier coefficients $\{ f_{\uv{\ben} n },\, f_{\uv{\cen} n }, \, f_{z n } \}$  of course only depend on $k$ and not on $\theta$. With $\mathcal{E}:=E_x+iE_y$ one has $
2(E_\uv{k}\uv{\ben}\pm E_\uv{\theta}\uv{\cen})= e^{i\theta}\mathcal{E}^* (\uv{\ben}\pm i \uv{\cen})+
e^{-i\theta}\mathcal{E} (\uv{\ben}\mp i \uv{\cen})$
 and therefore the  left-hand side of \eref{fLE} can be written as
\be{n1}
\frac{1}{2}e^{i\theta}\mathcal{E}^* 
\mtrx{ccc}
 {\uv{\ben} & \uv{\cen} & \uv{z} }
\mtrx{c} {\Ff+\Gf \\i\Gf-i\Ff \\ 0}+\textrm{c.c.}
\, , 
\ee
which contains only the $n=1$ Fourier component and the complex conjugate $n=-1$ component. This is going to imply that 
$f_{\uv{\ben} n },\, f_{\uv{\cen} n }, \, f_{z n } =0$ for $|n|\neq 1$. (Choosing a cartesian basis in \eref{decomp}, in contrast,  couples the equation for component $n$ with the components $n\pm N$.) This fact has some direct implications for the electric field induced contributions to the real space densities. For example,  the real space density of $z$-spins  $f_z^{(E)}$ is trivially zero. For $|N| \neq 1$ the in-plane components $f_x^{(E)}$ and $f_y^{(E)}$ also vanish trivially in real space, whereas for $|N|=1$ they can be nonzero. The real space spin Hall current $ \vec{j}^z $ can be nonzero for all $N$, whereas  the contribution to $\vec{j}^x$ and $\vec{j}^y$  vanishes trivially for all $N$.

On the right-hand side of \eref{inc1} we find
\be{inc2}
-\int_{\theta\pr}  K \DT \vec{f}|_{k\pr=k}^{(E)}=-
\sum_n 
e^{in\theta}
\left( 
\begin{array}{rcl}
& \uv{\ben} & \left( f_{\uv{\ben}n}    
 \int_{\theta\pr} K  (1-\cos N\DT \theta \cos n\DT \theta )
+i f_{\uv{\cen}n} \int_{\theta\pr}  K \sin N \DT \theta\sin n\DT\theta   \right)+ 
\nonumber \\
 +&\uv{\cen}  & 
  \left( f_{\uv{\cen}n}  
\int_{\theta\pr} K  (1-\cos N\DT \theta \cos n\DT \theta )
-i f_{\uv{\ben}n} \int_{\theta\pr}  K \sin N \DT \theta\sin n\DT\theta   \right)+ 
\nonumber \\
+& \uv{z}  & f_{zn} 
  \int_{\theta\pr} K (1- \cos n\DT \theta ) 
\end{array}
\right)\, .
\ee
Here, $\uv{\ben\pr}=\uv{\ben} \cos N\DT\theta-\uv{\cen}\sin N \DT\theta$ was used and terms odd in $\DT \theta$ were left out, using that $K$ is even in $\DT \theta$.
Including the precession term $2\ben (\uv{z}f_\uv{\cen}-\uv{\cen}f_z)$ 
the equation for the  $n=1$ Fourier components of $f^{(E)}$ becomes
\be{matrixb}
\frac{\mathcal{E}^*}{2}
\mtrx{c} {
\Ff+\Gf 
\\
i\Gf-i\Ff 
\\
 0} 
 =
-
\mtrx{ccc}{
\tcos^{-1} & i \tsin^{-1} & 0 \\
-i\tsin^{-1} & \tcos^{-1} & 2\ben \\
0 & -2\ben & \ttr^{-1}}
\mtrx{c}
{f_{\uv{\ben}1} \\
f_{\uv{\cen}1} \\
f_{z1}
}
\ee
where 
\be{}
\tcos^{-1} &:=&\int_{\theta\pr} K  (1-\cos N\DT \theta \cos \DT \theta )=K_0-(K_{N-1}+K_{N+1})/2 \nonumber \\
\tsin^{-1}&:=& \int_{\theta\pr}  K \sin N \DT \theta\sin \DT\theta =(K_{N-1}-K_{N+1})/2
\ee
(possibly negative) were introduced.

With the $\lambda$-dependent $\ben$, the matrix and hence the solution will be inhomogenous in $\lambda$.  For $|N|\neq 1$ the solution nonetheless goes to zero when $\lambda\rightarrow 0$. For point-like impurities ($\tcos^{-1}=\ttr^{-1}=K_0$ and $\tsin^{-1}=0$) we find   
\be{n3}
f_{z1}=i\frac{N\mathcal{E}^*\lambda^2 k \partial_\aen \FDf }{4k^2\lambda^2+K_0^2}
\Longrightarrow 
\vec{j}^z =\int_\vec{k} \vec{v}_0 f_z= \frac{N k_\mathrm{F}^2\lambda^2}{2\pi (4 k_\mathrm{F}^2\lambda^2+K_0^2)}(-E_y, E_x)\, 
\ee
for the real space spin current. 
As in ref.~\onlinecite{Shy} one recovers in  the {\it clean} limit $\ttr^{-1}=K_0\ll  k_\mathrm{F}\lambda $  a universal spin Hall conductivity $\sigma^\mathrm{SH}=\frac{N}{8\pi}$, whereas $\sigma_\textrm{SH}$ decreases to zero in the opposite ({\it dirty}) limit.

For the case $N=\pm 1$ one has for arbitrary impurity range that $\tcos= \pm\tsin=:2 \tau_{02}$. The determinant $4\ben^2 \tcos^{-1}+\ttr^{-1}(\tcos^{-2}-\tsin^{-2})$  of the matrix in \eref{matrixb} becomes  singular at $\ben = 0$. For $N=1$ we find the solution
\be{shysol}
f_{\uv{\ben}1} & = & -\mathcal{E}^* \left( \tau_{02}
(\Ff+\Gf)  +\frac{\Gf}{4 \ttr  \ben^2} \right)
  \nonumber \\
f_{\uv{\cen}1} & = & -
i\mathcal{E}^*\frac{ \Gf}{4\ttr \ben^2} 
\nonumber \\
f_{z1} & = & -
i\mathcal{E}^*\frac{ \Gf}{ 2 \ben} \, . 
\ee
The $z$-component of the real space spin density $\int_\vec{k}f_z$ must vanish trivially as noted after equation \eref{n1}. The in-plane spin components, resulting in a electric-field-induced polarization in real space
\be{sE}
\vec{s}=\int_\vec{k} (\uv{b}f_\uv{b}+\uv{c}f_\uv{c}) =-\uv{z}\times \vec{E} \int \der \aen\, \DS  \left( \tau_{02}
(\Ff+\Gf)  +\frac{\Gf}{2 \ttr  \ben^2} \right)\, .
\ee
Remember that the contribution from the equilibrium polarization is zero in real space.

According to \eref{FandG}, $\Gf=0$ for  $N=1$ for arbitrary dispersion and arbitrary range of impurities according. Thus,  $f_z=0$ for a Rashba coupling, implying a zero spin Hall current. A similar analysis applies to $N=-1$, \eg for the linear Dresselhaus coupling,  since a  model of winding $-N$ is related to one of winding $N$ through reflection (say along the $x$-axis).

If for the Rashba coupling there was a nonzero spin Hall effect, \ie $\Gf\neq0$, then $f_z$ and consequently the spin Hall current would be indpendent of $\lambda$ (since in general, $\Gf\propto \lambda$). Furthermore, the in-plane components $f_x$ and $f_y$  of the polarization would diverge as $\lambda^{-1}$ as $\lambda\rightarrow 0$, \ie one would not recover an unpolarized distribution when sending the spin-orbit coupling to zero. However, in the derivation of the Boltzmann equation there were only assumptions of  $\lambda$ and $\vec{f}$ being small enough, no assumptions of them not being too small. Therefore, to the extent that a diverging polarization in such a case is an unthinkable result, the vanishing of the SHE for the Rashba case is a natural implication.

\section{General arguments for a vanishing SHE}\label{s:ga}
\noindent
The vanishing of  the SHE for the Rashba case has been found by numerous previous studies (see the introduction). To the knowledge of the author  it has not been related to the finiteness of the in-plane polarization like done in the argument above. In linear response studies general arguments have been given in the case of quadratic dispersion $\aen\propto k^2$ \cite{Dim, ChaLos}, where it is noted from $[\sigma_y,\lambda (\vec{p}\times\vec{\sigma})_z ]=2i\lambda p_y\sigma_z$  that for Heisenberg operators 
\be{}
\frac{\der}{\der t} \hat{\sigma}_y =-i[\hat{\sigma}_y,\hat{H}+\hat{V}_\textrm{imp}] =2\lambda\hat{p}_y\hat{\sigma}_z=2 \lambda m \hat{j}_y^z 
\ee
for non-magnetic impurities. The steady state condition $\langle \frac{\der}{\der t} \hat{\sigma}_y\rangle=0$  forces the spin Hall  current $\langle \hat{j}_y^z\rangle$ to be zero. A similar argument applies to $ \hat{j}_x^z$.

 The analog to this argument in the  Boltzmann approach can be established for the real space densities. In phase space,  $(\vec{s}\times\vec{b})^y=b_xs^z=j_y^z b_x/v_{0y}$. Only  for $N=\pm 1 $ and for $\ben/\vo$ independent of $k$  is it possible for  $b_x/v_{0y}$ to be a constant, here $\mp \lambda m$. According to \eref{lhs2} one has
$\partial_t s^y=-2(\vec{s}\times\vec{b})^y-\vec{E}\cdot \partial_\vec{k}\vec{s}^y+\vec{\mathcal{J}}$ in the uniform case. In the integration over momentum the last two terms vanish, resulting in  $\partial_t s^y=\pm 2\lambda m j_y^z$ {\it  in real space} for $N=\pm 1$. Likewise, $\partial_t s^x=- 2\lambda m j_x^z$. In a steady state, $\partial_t \vec{s}=0$ implies $\vec{j}^z=0$. 

In phase space, on the other hand, the steady state condition $0=\partial_t s^y =-2(\vec{s}\times\vec{b})^y-\vec{E}\cdot \partial_\vec{k}\vec{s}^y \ldots $ does not imply a vanishing $j_y^z$. We could have had  $G\neq 0$ as long as $\int \der \aen\,  \DS v_0 G/b=0$ guaranteed the vanishing in real space. However, in the previous section we find that  $\vec{j}^z=0$ also in phase space.

\section{Assumption of linear response in the spin-orbit coupling}\label{s:assumption}
\noindent
We are now going to study the implications of a basic assumption, namely that the steady-state response of a spin-orbit coupled system in an electric field is perturbative in the small parameter $\lambda$ ($ v_\mathrm{F}$ is kept constant), analogously to the usual assumption  of linear response in the electric field strength $E$. For the latter assumption it means that one can expand the solution $f^\textrm{eq}+f^{(E)}+f^{(E^2)}+\ldots $ in {\it nonnegative} powers in the electric field and solve the equation iteratively by solving equations that are homogenous in orders of $E$, see \eg equation~\eref{inc1}. 
With the assumption now to be studied, the same is expected to apply also for the spin-orbit coupling parameter $\lambda$ ($v_\mathrm{F}$ is kept constant), wherefore the solution can be expanded in powers of $E$ and $\lambda$ 
\be{}
f=\sum_{m,n\geq 0} f^{(\lambda^m, E^n)}\, 
\ee 
assuming that the solution is analytic both in $\lambda$ and $E$. 

The equations are going to be solved order by order in both parameters as illustrated in \eref{LE}. In the expansion of the Boltzmann equation in a small spin-orbit coupling, there is in this case no ambiguity about which terms to include in a given equation. This should be contrasted with the previous section, where we chose  to include the precession term $2\vec{\ben}\times\vec{f}$ ---a term of order $\ordo{\lambda^2}$---though we discarded other terms of the same order, for example in \eref{jlin}. The reproduced  results seem to be based on hand-picking terms with physical insight.


\section{Boltzmann equation without the precession term}\label{s:bewa}
\noindent
Due to the assumption in sec. \ref{s:assumption}, not only the equation but also the solution is linearized in the spin-orbit coupling. Together with the habitual linearization in the electric field
this means that we consider 
\be{expf}
f=f^{(0)}+f^{(\lambda)}+f^{(E)}+f^{(\lambda,E)}
\ee   
where the superscripts denote the order in $\lambda$ and $E$, respectively (\eg $f^{(E)}\propto \lambda^0E^1$). The equilibrium distribution is $f^\mathrm{eq}=f^{(0)}+f^{(\lambda)}$, where $f^{(0)}=\idm \FDf$. 

The left-hand side of the Boltzmann equation \eref{lhs1}  is in the static, uniform case 
\be{}
 \frac{\der f}{\der t}=\vec{E}\cdot\partial_\vec{k} f+
2\vec{\sigma}\cdot \overbrace{\vec{f}\times\vec{\ben}   }^{\ordo{\lambda^2}} 
\ee
where the precession term must be neglected to order $\lambda$. However, leaving out the precession term leads to some formal trouble in the case $N=\pm 1$. The derivative $\vec{E}\cdot\partial_\vec{k}=e^{i\theta}(E_x-iE_y)(\partial_k+ik^{-1}\partial_\theta) +\textrm{c.c.} $ comes with a winding number $\pm 1$. For a  spin-orbit coupled system  the equilibrium polarization $\vec{f}^\textrm{eq}=\vec{f} (E=0)$ has a also winding number, here $N$. Thus, the combination $\vec{E}\cdot\partial_\vec{k}f$ comes with terms of winding  $N\pm1$. For $N=1$ (\eg Rashba) or $N=-1$ (\eg linear Dresselhaus) the left hand side  therefore contains terms without angular dependence, as seen in \eref{angindep}. Such terms cannot be matched by the collision integral, essentially because an equation like $1=\int \der \theta\pr\, (f(\theta)-f(\theta\pr))$ has no solution. (A collision integral cannot be a source/drain.) 

This lack of solution might be related to the discarding of  the principal part terms $X$ in \eref{jcb}, but we have not been able to investigate this.  
As a simple remedy, we introduce instead a small spin relaxation term, which could come from spin relaxation processes not related to the spin-orbit coupling. The Boltzmann equation then reads 
\be{eboltz}
\vec{\sigma}\cdot \taus^{-1}\vec{f} +\vec{E}\cdot \partial_\vec{k} f  
&=& -\ksum{\vec{k\pr}} \Wkk\,\, \left[
 \delta (\DT\aen )
\DT f + 
\delta\pr (\DT \aen) \vec{\sigma}\cdot \DT \vec{\ben}  \DT f_0
  \right]
\ee
where it is assumed that $\taus^{-1}$ is much smaller than $\ttr^{-1}$, is independent of $\lambda$ and is a number though in general it could be a matrix. 

We now solve the Boltzmann equation \eref{eboltz} order by order in $\lambda$ and $E$ as shown in equation~\eref{LE}
\be{LE}
\begin{array}{lrcl}
{\large \lambda^0 E^0: }\hspace{1cm}
&0 & = & -\ksum{\vec{k\pr}} \Wkk\delta(\DT \aen) \DT f^{(0)} \\
{\large \lambda^1 E^0: }
 &\vec{\sigma}\cdot  \taus^{-1}\vec{f}^{(\lambda)}   & = & -\ksum{\vec{k\pr}} W_{\vec{k}\vec{k\pr}}
\left[ \delta (\DT \aen) \DT f^{(\lambda)} +\delta\pr (\DT \aen) 
\vec{\sigma}\cdot \DT \vec{\ben}  
\DT f_0^{(0)}  \right] 
\\ 
{\large \lambda^0 E^1: }
 & \vec{E}\cdot\partial_\vec{k} f^{(0)} &=&-\ksum{\vec{k\pr}} W_{\vec{k}\vec{k\pr}}
\delta (\DT \aen) \DT f^{(E)} 
\\
{\large \lambda^1 E^1: }
 & 
 \vec{\sigma}\cdot  \taus^{-1}\vec{f}^{(\lambda, E)} + \vec{E}\cdot\partial_\vec{k} f^{(\lambda)} &=&  
  -\ksum{\vec{k\pr}} W_{\vec{k}\vec{k\pr}}
\left[ \delta (\DT \aen) \DT f^{(\lambda,E)} +\delta\pr (\DT \aen) 
\vec{\sigma}\cdot \DT \vec{\ben} 
\DT f_0^{(E)}  \right]     
\end{array}\, .
\ee  
Note that terms known to be zero were left out.\footnote{
For example $\vec{f}^{(E)}=0$ since the electric field alone leads to no polarization, therefore $\taus^{-1}f$ could be left out in the equations of order  $\lambda^0E^x$.} We are out to find $f^{(\lambda,E)}$, which is of order $\lambda^1E^1$ and therefore the first contribution that incorporates the combined effects of an electric field and spin-orbit coupling.\footnote{
Note that $f^{(\lambda^2)}$ or $f^{(E^2)}$ cannot contribute to the spin Hall effect.  Therefore we do not need consider them, though one of them need not be subleading  to $f^{(\lambda,E)}$. 
}  
However, to solve the $\lambda^1E^1$ equation one needs to solve the previous equations to find out $f^{(\lambda)}$ and $f^{(E)}$. 

The $\lambda^0E^0$ equation is consistent with $f^{(0)}=\idm \FDf$ from \eref{equil2}.  The  spin-relaxation term in the $\lambda^1E^0$ equation decreases the equilibrium  polarization  in \eref{equil2} into $f^{(\lambda)}=
\tstr\vec{\sigma}\cdot \vec{\ben}\partial_\aen \FDf$ with $\tstr:=(1+\taus^{-1}\ttr)^{-1} \approx 1$. The $\lambda^0E^1$ equation is  solved by $f^{(E)} = -\idm \ttr E_\uv{k} \partial_k \FDf$. 
So finally at the $\lambda^1 E^1$ equation one knows already the terms $\vec{E}\cdot\partial_\vec{k} f^{(\lambda)}$  and $\ksum{\vec{k\pr}} W_{\vec{k}\vec{k\pr}} \delta\pr (\DT \aen) 
\DT \vec{\ben} \cdot\vec{\sigma}\DT f_0^{(E)}$. Collecting these contributions on the left-hand side we get an equation of the same form as \eref{fLE}, but with the spin precession term  replaced by the spin relaxation term, and with $\Ff$ and $\Gf$ modified due to the factor $\tstr$ in $\vec{E}\cdot \partial_\vec{k} f^{(\lambda)}$. The  equation for the $n=1$ Fourier coefficients turns into
 \be{f1eq}
\frac{\mathcal{E}^*}{2}
\mtrx{c} {
\Ff+\Gf 
\\
i\Gf-i\Ff 
\\
 0} 
 =
-
\mtrx{ccc}{
\taus^{-1}+\tcos^{-1} & i \tsin^{-1} & 0 \\
-i\tsin^{-1} & \taus^{-1}+\tcos^{-1} & 0\\
0 & 0 & \taus^{-1}+\ttr^{-1}}
\mtrx{c}
{f_{\uv{\ben}1} \\
f_{\uv{\cen}1} \\
f_{z1}
}
\ee
 For $N=\pm1$ one has $\tcos=\pm\tsin$ for a general impurity potential, leading the determinant of the matrix to be  zero unless $\taus^{-1}\neq 0$. (With $\taus^{-1}=0$ there is either no solution when $\Gf\neq 0$ or multiple solutions when $\Gf =0$.) 
For $|N|\neq 1$ one has $\tcos\neq \tsin$ for realisitc impurity potentials, wherefore $\taus^{-1}\neq 0$ is not needed. 

Important to retain is that the matrix does not depend on $\lambda$. The polarization $\vec{f}^{(\lambda,E)}$ is therefore proportional to $\lambda$. Clear is also that $f_\uv{z}=0$ and the  spin Hall current $\vec{j}^z$ are zero---for an arbitrary spin-orbit coupling. 

For $N=\pm 1$ we can adopt the derivation in sec.~\ref{s:ga} to show that $\partial_t s_y=\pm m\lambda j^z_y -\taus^{-1}s_y$ in real space. For our steady state case we therefore find $s^y \propto j^z_y= 0$ in real space for $N=\pm1$.  Likewise, $s^x=0$. For $|N|\neq 1$ the real space polarization is trivially zero (see  under equation~\eref{n1}).  Summarizing we have that for no $N$ in the perturbative case does the electric field alter any real space densities to lowest order in $\lambda$. The only nonzero real space densities are the equilibrium spin currents $\vec{j}^x$ and $\vec{j}^y$ for $|N|$=1.  From  $f^\mathrm{eq}=\FDf+\tstr \vec{\sigma} \cdot \vec{\ben}\partial_\aen \FDf$  for $N=\pm1$ we find $
j_y^x=\mp j_x^y=\frac{1}{4\pi}(\tstr-1) \ben_\mathrm{F}k_\mathrm{F}  
$.   

Table~\ref{t:table} gives a summary and offers a comparison of the different cases consider in this paper.  In this paper we do not consider the case $N$ even, but we note that for $|N|=2$  
an electric-field-induced contribution to the real space spin currents $\vec{j}^x$ and $\vec{j}^y$  is not forced  to be zero by simple angular considerations. (This is analogous to  $\vec{s}^{x(E)}$ and $\vec{s}^{y(E)}$  being allowed to be nonzero only for $|N|=1$.) The case $N=2$ is relevant for pseudospin currents for one valley in bilayer graphene. Note that this is not the SHE pseudospin current $\vec{j}^z$ discussed in ref.~\onlinecite{FujJalTan}. Actually, with the definition \eref{spincurr} a nonzero (pseudo)spin current $\vec{j}^z$ is not possible when $\aen=0$, the latter being the case in single and double layer graphene.  

\begin{table}[htdp]
\begin{center}
\begin{tabular}{c|c|cccc} 
 \multicolumn{6}{c}{Nonperturbative} \\
   \multicolumn{2}{c}{}    & $s^{x,y}$ &$ s^z$  & $j^{x,y}$ & $j^z$  \\
 \hline
$ |N|=1$  & $\feq $    & \nz  /  0                      & 0/ 0                 & \nz   /  \nz${}^{1)}$ &  0/ 0 \\
                 & $f^{(E)}$ & \nz  /  \nz${}^{2)}$   & 0${}^{3)}$/ 0  & \nz/ 0                         &  0${}^{3)}$/ 0${}^{4)}$ \\
   \hline
 $ |N|\neq 1$  &  $\feq$     & \nz  /  0   & 0/ 0      & \nz   /  0             &  0/ 0 \\
                         & $f^{(E)}$ & \nz  /  0   &  \nz/ 0  & \nz/ 0${}^{5)}$ &  \nz/ \nz \\
  \multicolumn{6}{c}{Perturbative} \\
    
  \multicolumn{2}{c}{}   & $s^{x,y}$ &$ s^z$  & $j^{x,y}$ & $j^z$  \\
  \hline
$ |N|=1$  & $\feq $    & \nz  /  0                      & 0/ 0                 & \nz   /  \nz${}^{6)}$ &  0/ 0 \\
                 & $f^{(E)}$ & \nz  /  0${}^{6)}$   & 0/ 0                 &      \nz / 0                         &  0/ 0\\
   \hline
 $ |N|\neq 1$  &  $\feq$     & \nz  /  0   & 0/ 0      & \nz   /  0             &  0/ 0 \\
                         & $f^{(E)}$ & \nz  /  0   &  0/ 0  & \nz/ 0${}^{5)}$ &  0/ 0          \\
\end{tabular}
\end{center}
\caption{Summary of spin densities and spin currents in the nonperturbative and the perturbative cases for the cases $N=\pm1$ and for other odd $N$. The entries present the (phase space)/(real space) quantities, respectively. \nz{} stands for a (typically) nonzero result whereas uncommented zeros stand for trivially vanishing components (\eg due to angular considerations). Nontrivial cases are commented. 
1) $j_x^y, j_y^x =\ordo{\lambda^3} $, see under \eref{equil}. 
2) See \eref{sE}.
3) Since $\Gf=0$.
4) General arguments in sec.~\ref{s:ga}. 
5) Vanishing for $N$ odd. Could, however, be nonzero for  $|N|=2$.
6) See end of sec.~\ref{s:bewa}.  Relies on the introduced spin relaxation term.  Here $j_x^y, j_y^x =\ordo{\lambda} $
} 
\label{t:table}
\end{table}%

\section{Electric field induced corrections to the collision integral}
\label{s:sh}
\noindent
The precession term seemed so far like the only term that could involve the  $f_z$  component and lead to nonzero spin current $\vec{j}^z$. In this section we are going to see that the electric field \vec{E} modifies the collision integral in a way that involves the $f_z$ component. However, the corrections turns out to be of order $\lambda^2$. 

In trying to incorporate in our Boltzmann equation all terms present to first order in our parameters we so far left out terms by  gradient expanding the right-hand side of the Dyson equation \eref{gkeq} for $\GK$ only to zeroth order and not to first order in the electric field. First order corrections have been accounted for example in the case of electron-phonon renormalization of the ac conductivity  \cite{Hol} (see also  \cite{RamSmi}) and have also been discussed in the SHE and AHE literature (see \eg refs.~\onlinecite{Sug} or \onlinecite{Kov}).  However, a derivative such as $
2 \vec{E}\cdot\partial_\vec{k} \sproj_{\uv{\ben}s} = s k^{-1}E_\uv{\theta}\partial_\theta \uv{\ben}\cdot \vec{\sigma}$, which occurs in the derivation of these corrections, gives a vector that remains in the plane.  The collision integral becomes more complicated but does not involve the $f_z$ component. 

The contribution discussed in this paper comes about in a slightly subtler way, and has to the knowledge of the author not been discussed previously. It comes from gradient expanding the equation of motion also for the {\it retarded} Green's function $\GoR$, which results in the correction $\delta\GoR$ given by \eref{gor2}. With $B_z=0$ one obtains
\be{deltaJ}
\delta \mathcal{J} &=&
-  \ki \frac{\Wkk}{2\pi} \int \frac{\der \omega}{2 \pi} (\delta \GoR_\vec{k} \DT f \GoA_\vec{k\pr}+
 \GoR_\vec{k} \DT f \delta\GoA_\vec{k\pr} + \GoR_\vec{k\pr} \DT f \delta\GoA_\vec{k} + \delta\GoR_\vec{k\pr} \DT f \GoA_\vec{k})=
 \nonumber \\
 &=& - \sum_{ss\pr}\ki \frac{\Wkk}{8}
 \left[ 
 \begin{array}{rl}
 ( & \sigma_z \DT \sproj\pr+\sproj\pr\DT f \sigma_z )
 E_\uv{\theta} (\partial_\aen -s\ben^{-1})\delta (\DT \erg)+
 \\
 + & 
( \sigma_z \DT \sproj+\sproj\DT f \sigma_z )
 E_\uv{\theta\pr} (\partial_{\aen\pr} -s\pr{\ben\pr}^{-1})\delta (\DT \erg) 
 \end{array} 
 \right] =
 \nonumber \\
&=& \vec{\sigma}\cdot \frac{1}{2} \ki \delta^{\prime\prime} (\DT \aen)
\left ( \Wkk 
\left[\uv{z} \Mkk\cdot \DT \vec{f}+\Mkk \DT f_z    
\right] 
\right)
\ee
where for the last line the expansion \eref{deltexp} was used and where
\be{}
\Mkk (\vec{k},\vec{k\pr}) :=  k^{-1}E_\uv{\theta}\vec{\ben\pr }+ {k\pr}^{-1}E_\uv{\theta\pr}\vec{\ben } \, .
\ee
Note that $\delta \mathcal{J}_0=0$, i.e. there is no contribution to the charge part of the equation, only to the polarization part. Note particularly that $\delta\mathcal{J}_z\neq 0$, which would give a nontrivial equation for $f_z$.
However, since $\Mkk \propto \lambda E$ and $\vec{f}\propto \lambda$, this correction contributes to order $\lambda^2E$ as announced, wherefore the result $f_z=0$ in last section is not changed to lowest order.

One can ask if this correction could change the result $f_z=0$ for the Rashba case in the non-perturbative scenario of sec.~\ref{s:pt} where homogenity in $\lambda$ was not an issue.  The correction would then enter as an imaginary element of order $\lambda^2 E$ replacing the zero in the bottom of the vector with $\Ff$ and $\Gf$ in \eref{matrixb}. However, since $\tcos^{-2}-\tsin^{-2}=0$ for $N=1$, the inverse matrix has a zero $zz$-element. Therefore, the correction cannot contribute to $f_{z1}$ in the Rashba case. In this aspect the contributions discussed in the beginning of this section, on the other hand, could  contribute, but we have not investigated them in a systematic way. The real space spin current $\vec{j}^z$ would in any case remain zero, at least for a quadratic dispersion,  due to the arguments in sec.~\ref{s:ga}.

\section{Conclusions} 
\noindent
This paper studied the intrinsic contribution to the spin Hall current in a spin-orbit coupled 2DEG by deriving a Boltzmann equation in the Keldysh formalism and solving it in the uniform steady state case. The vector $\vec{\ben}$ determining the spin-orbit coupling was assumed to be of the form  $\vec{\ben}=\ben(k)\uv{\ben}(\theta)$ with $\ben =\lambda k$ and $\hat{\ben}_x+i\hat{\ben}_y \propto e^{i\theta_0 +iN\theta}$. We reproduced the common result that spin Hall effect vanishes for $N=\pm1$ (\eg for a Rashba coupling) but not for other $N$. We were able to give a new perspective on this vanishing by pointing out that a nonzero result leads the in-plane components of the polarization to diverge when $\lambda\rightarrow 0$.

The mentioned treatment does not assume the response to be perturbative in $\lambda$.
We therefore found it interesting to study the implications of assuming the response to be perturbative not only in the electric field but also in $\lambda$. The precession term---previously the prerequisite for the spin Hall current---must then be left out to first order in spin-orbit splitting. The out-of-plane polarization $f_z$ becomes trivially zero and there seems to be zero SHE for any winding. We saw also that all other real space densities have zero electric field induced contributions.

Leaving out the spin-precession term gives a Boltzmann equation for in-plane polarization $(f_x,f_y)$  that is unsolvable for $N= \pm 1$. The unsolvability might be related to the left out principal parts, the inclusion of which would have been beyond the scope of the present study. As an ad hoc remedy the precession term got replaced by a small spin relaxation term. This could  suggest that if the response is perturbative in $\lambda$, then non-magnetic impurities are not enough for the existence of a steady state solution for the polarization.

To cover all contributions to first order in electric field, in spin-orbit splitting and in impurity strength and concentration, we considered corrections to the collision integral that come from going to first order in electric field in the gradient expansion of the self-energy side of the Dyson equation. One of the corrections, to our knowledge not discussed before, actually involves the $f_z$ component. However, this contribution is of order $\lambda^2$. Thus, the vanishing of the spin Hall current to lowest order in $\lambda$ in the perturbative case is not changed by these corrections. 

Finally, the paper includes a detailed discussion of why a relaxation time approximation fails
 and a comment on pseudospin currents in bilayer graphene.

{\it Aknowledgements}.
The author wishes to aknowledge discussions with M.~L\"uffe, D.~Culcer, F.~Gethmann,  A.~G.~Mal'shukov, K.~Morawetz, T.~Nunner, P.~Schwab, G.~Vignale, F.~von Oppen and R.~Winkler. This work was supported by the Swedish Research Council.  The author also wishes to acknowledge a visitor grant of the Max Planck Institute for the Physics of Complex Systems.

\appendix
\section{Left hand side of eq. \eref{fLE}} \label{a:lhs} 
\noindent
For $\ben=\lambda k$ and $\hat{\ben}_x+i\hat{\ben}_y=e^{i\theta_0+iN\theta}$ and $\vec{f}=\tstr \vec{\ben}\partial_\aen \FDf $ one has
\be{fL}
\vec{E}\cdot \partial_\vec{k} \vec{f}
=
(E_\uv{k}\partial_k +\frac{1}{k}E_\uv{\theta}\partial_\theta) \tstr \lambda k \uv{\ben} 
\partial_\aen \FDf 
=
 \tstr\lambda (E_\uv{k}\uv{\ben} +N E_\uv{\theta}\uv{\cen})\,\partial_\aen \FDf+
\tstr \lambda k\, E_\uv{k}\uv{\ben}\,\partial_k\partial_\aen \FDf \, .
\ee
The term $\ksum{\vec{k\pr}} W_{\vec{k}\vec{k\pr}} \delta\pr (\DT \aen) 
\DT \vec{\ben} \cdot\vec{\sigma}\DT f_0^{(E)}$  is here for brevity only evaluated for a point-like impurity potential, \ie $W_{\vec{k}\vec{k\pr}}=W$ constant, and constant density of states $\DS(\aen)=m/2\pi$. Hence 
\be{}
\ttr^{-1}= \ksum{\vec{k\pr}} \delta (\DT \aen) W_{\vec{k}\vec{k\pr}} (1-\cos (\DT \theta)) 
 =\DS W
 \ee 
With  
$\delta\pr (\DT \aen)=\partial_\aen\delta (\DT \aen)=-\partial_{\aen\pr}\delta (\DT\aen)$
a partial integration gives 
\be{fE}
\ksum{\vec{k\pr}} W \delta\pr (\DT \aen) 
\DT \vec{\ben} \DT f_0^{(E)} &=&
\int \der \aen\pr \frac{\der \theta\pr}{2\pi} \delta(\DT \aen) \partial_{\aen\pr}
(\DS W \DT \vec{\ben} \DT f_0^{(E)})= \nonumber\\
&=&\lambda \int \der \aen\pr \delta(\DT \aen)
\left[ 
 k\partial_\aen\partial_k\FDf \int  \frac{\der \theta\pr}{2\pi}  E_\uv{k\pr} \DT \uv{\ben}+
 \partial_{\aen\pr}(k\pr) \partial_k\FDf  \int  \frac{\der \theta\pr}{2\pi} \uv{\ben\pr} \DT E_\uv{k\pr}   \right] = \nonumber \\
&=& -\delta_{|N|,1}\frac{\lambda}{2}(E_\uv{k}\uv{\ben}+N E_\uv{\theta} \uv{\cen} )
  (k\partial_\aen\partial_k \FDf+\partial_\aen \FDf)= \nonumber \\
& =&   
    -\delta_{|N|,1}\frac{\lambda}{2}(E_\uv{k}\uv{\ben}+N E_\uv{\theta} \uv{\cen} )
  (k\partial_k\partial_\aen \FDf+\zeta\partial_\aen \FDf) 
\ee   
where it was used that $\uv{\ben\pr}=\uv{\ben} \cos N\DT\theta-\uv{\cen}\sin N \DT\theta$ and 
$\uv{k\pr}=\uv{k} \cos \DT\theta-\uv{\theta}\sin\DT\theta$.
In the last line it was used that $\partial_\aen\partial_k=
 \partial_\aen( \frac{\der \aen }{\der k}\partial_\aen )=
 (\partial_\aen v_0) \partial_\aen +v_0\partial_\aen^2= \frac{\zeta-1}{k}\partial_\aen+\partial_k\partial_\aen$ for  $v_0\propto k^{\zeta-1}$. For a non-constant $\DS$ (\ie for $\zeta\neq 2$) the result in \eref{fE} is modified. Note also that it is only for point-like impurities that the contribution \eref{fE} vanishes for $|N|\neq  1$.  
 
 Adding up \eref{fL} and \eref{fE} one obtains for $N=1$ that $\Ff=\frac{1}{2}\lambda(\zeta-2)\partial_\aen\FDf+ \frac{1}{2}\lambda \tstr k \partial_k\partial_\aen \FDf$ and $\Gf=\lambda (\tstr-1)(\partial_\aen \FDf+\frac{1}{2}k\partial_k\partial_\aen\FDf)$. 
For $|N|\neq 1$ the contribution \eref{fE} vanishes and one gets for example $\Gf-\Ff=N \lambda \tstr \partial_\aen \FDf$, needed for the result \eref{n3}.

  
\section{Failure of a relaxation time approximation}\label{a:rta}
\noindent
A useful approximation for the collision integral found in standard applications of the Boltzmann equation is the {\it relaxation time approximation} (RTA)
\be{rta}
\JC[f_\vec{k}] = - \int_\vec{k\pr}   \delta (\DT \epsilon) W_{\vec{k} \vec{k\pr}} (f_\vec{k}- f_\vec{k\pr})  \longrightarrow 
-\frac{
\delta f_\vec{k}
}{\tau}\, ,
\ee
where the relaxation time $\tau$  depends only on the absolute value $k$ of the momentum. 
It expresses that the role of the collision integral is to relax the deviation from equilibrium
$\delta f_\vec{k}:=f_\vec{k}-\feq_\vec{k}$. For $\Wkk$ angularly independent (\ie momentum independent) it can be derived from the collision integral to the left in \eref{rta}. Thus RTA should be a good approximation for weakly momentum dependent impurity potentials. 

 A prerequisite for the RTA  is that it is consistent with the conserved quantities. In particular, the momentum integral of the Boltzmann equation should give the continuity  equation $\partial_t n +\partial_\vec{x}\cdot \vec{j}=0$ expressing the  conservation of particle number.  The right hand side (given $\Wkk=W_{\vec{k\pr}\vec{k}}$) indeed vanishes identically $\int_\vec{k} \JC[f_\vec{k}]=0$, expressing that the collision term  cannot in real space act as a source (or drain) of particles. For the  RTA the particle conservation is not automatic, since $\int_\vec{k} \delta f_\vec{k}/\tau$ would be nonzero if $\delta f$ contained an angularly independent component. However, in typical applications the deviation $\delta f=\sum_n \delta  f_n \exp (in\theta)$ contains to lowest order only angularly dependent terms. 
 
The  RTA can be implemented also in the case of spin in the same way as in \eref{rta}, or more generally by letting $\tau^{-1}$ be a matrix acting on $\delta \vec{f}$ to allow for different relaxation times for different spin components of $\vec{f}$.  
The RTA has been useful for example for the derivation of the Dyakonov-Perel spin relaxation mechanism (DPSR) \cite{DyaPer2} (see also refs.~\onlinecite{AveGol} and \onlinecite{GlaIvc}), found by deriving for initially polarized electrons the Bloch equations  from the uniform, time-dependent Boltzmann equation  for a Rashba-type spin-orbit interaction. In that problem the electric field is absent.

The general collision integral \eref{jc} still satisfies $\int_\vec{k} \JC=0$ for all components, expressing that not only particle number but also spin is conserved in collisions with {\it nonmagnetic} impurities.  Thus, a sensible RTA must still satisfy $\int_\vec{k} \delta f_\vec{k}/\tau=0$.  That is to say $\delta f$ cannot contain an angularly constant term.  In the derivation of the DPSR this is satisfied although $n=0$ components are present and essential.   The RTA is  only needed in  the part of the Boltzmann equation that is of first power in $\lambda$, and $f^{(\lambda)}$ is by simple inspection seen to only contain the angular components  $n=\pm1$.  

We note that in the SHE setup studied in this paper the situation is very different compared to the DPSR problem.  It turns out that a RTA always implies a nonzero SHE. For $N=\pm 1$  we therefore do not even qualitatively  reproduce the correct spin current. For  $|N|\neq1$  the SHE is on the other hand nonzero and can be captured by a RTA. In the $N=\pm1$ case the nonzero SHE is intimately related to the RTA failing to conserve spin. With the simple RTA $\JC[\vec{f}]\rightarrow -\delta \vec{f}/\tau$ the real space equations for the Rashba/Dresselhaus case in section~\ref{s:ga} get  modified to 
\be{rserta}
\partial_t s^y= \pm 2m\lambda j_y^z-\delta s^y/\tau\, .
\ee 
A steady state does no longer imply $j_y^z=0$ but instead $j_y^z=\pm \delta s^y/2m\lambda \tau$. Thus, if a polarization in real space is induced by  the electric field, then there is a nonzero SHE. The result in \eref{sE} shows that such a polarization {\it is} actually induced. Thus, the collision integral transformed into RTA form would enter as a nonzero spin source, simultaneously allowing for a nonzero SHE. Therefore, in contrast to the DPSR problem with a Rashba coupling, the RTA fails  when applied to the SHE problem  with a Rashba coupling. 

For $N=\pm 1$ in the perturbative case, a RTA and equation  \eref{rserta} do not lead to the same contradiction  since $s^x=s^y=0$ in real space (see table~\ref{t:table}). In the rest of the section we give some further details on the nonperturbative case.   

 In the SHE problem we look at  a steady state situation  and the expansion is done around the equilibrium distribution instead of an initial polarization as in the DPSR problem.  
 The Boltzmann equation is expanded in powers of  $E$ rather than in powers of $\lambda$. The lowest order deviation $\vec{f}^{(E)}$ (from the equilibrium polarization $\vec{f}^\textrm{eq}$) will now contain angularly independent terms, causing the non-vanishing of  $\int_\vec{k} \vec{f}^{(E)}$ in \eref{sE}. Contrast with ordinary steady problems with an electric field,  where constant terms are not present in
$f^{(E)}\sim \tau \vec{E}\cdot \vec{v} \partial_\epsilon\feq  $ since $\feq $ is independent of $\theta$ and   $ \vec{E}\cdot \vec{v}$ comes with angular components $n=\pm 1$. In the SHE problem on the other hand $\vec{f}^\textrm{eq} \propto \vec{\ben}$ comes with angular components $\pm N$, and in the Rashba/Dresselhaus case in particular $N=\pm1 $.  In the Rashba case $f^{(E)}$ can therefore contain angular components $n=\pm 1\pm 1=-2,0,2$ in {\it the cartesian basis}. (In the rotating basis given by $\{\uv{\ben}, \uv{\cen}, \uv{z}\} $ used in this paper it translates into the Fourier components $n=\pm1$. ) Particularly, the presence of the $n=0$ term means that a RTA would be inconsistent with the impurity scattering having to conserve spin. 

The other side of the coin---as displayed in \eref{rserta}---is that the RTA  allows for $f_{z1}\neq 0$ and consequently a nonzero SHE.  The analog of equation \eref{matrixb} is with a general matrix $\tau^{-1}$ with constant elements is given by  
 \be{matrixeqrta}
\frac{\mathcal{E}^*}{2}
\mtrx{c} {
\partial_\vec{k} \feq_\uv{\ben}
\\
i\frac{1}{k}\feq_\uv{\ben}     
\\
 0} 
 =
-
\mtrx{ccc}{
\tau_{bb}^{-1} & \tau_{bc}^{-1}   &  \tau_{bz}^{-1}  \\
\tau_{cb}^{-1}  & \tau_{cc}^{-1}  & 2\ben+\tau_{cz}^{-1}  \\
\tau_{zb}^{-1}  & -2\ben+\tau_{zc} ^{-1} & \tau_{zz}^{-1} 
}
\mtrx{c}
{f_{\uv{\ben}1} \\
f_{\uv{\cen}1} \\
f_{z1}
}\, .
\ee
Note that now the left hand side derives only from the term $\vec{E}\cdot \partial_\vec{k} \vec{f}^\textrm{eq}$, whereas the left hand side of \eref{matrixb} had an additional contribution from the term $ \ki W_{\vec{k}\vec{k\pr}}\delta\pr (\DT \aen) 
 \DT \vec{\ben}  
\DT f_0^{(E)}$ as seen in \eref{inc1}. {\it This latter term cannot be captured by a RTA. } However, this term was essential for giving a left hand side in \eref{matrixb} the crucial structure that all the components of the vector are proportional to each other, \ie linearly dependent, as followed with $\Gf=0$. This structure makes it possible to  find a solution $f_{z1}=0$ for suitable matrix elements in $\tau$. (In \eref{matrixb} we had $\tcos=\pm \tsin$.)
In the vector of  the left hand side of \eref{matrixeqrta}, on the other hand, the components $ \partial_\vec{k} \feq_\uv{\ben}$ and $k^{-1}\feq_\uv{\ben} $  are linearly independent. There is  no natural  choice of matrix elements of $\tau^{-1}$ that with such a left hand side can result in the cancellations needed to make $f_{z1}$ vanish.  (The elements of $\tau^{-1}$ are assumed to be at the most weakly dependent on $k$ and should certainly not contain factors of Fermi-Dirac distributions.) For the same reasons also a simple collision integral like $\JC=\ki \Wkk \delta (\Delta \aen) \Delta f$  fails to reproduce the vanishing SHE  because it gives a Boltzmann equation with the same left hand side as in \eref{matrixeqrta}.

For the case $|N|\neq 1$ the term  $ \ki W_{\vec{k}\vec{k\pr}}\delta\pr (\DT \aen) 
 \DT \vec{\ben}  
\DT f_0^{(E)}$ vanishes for $W$ constant (corresponding to $\tau$ constant), as seen in appendix~\ref{a:lhs}. The collision integral enters the Boltzmann equation only with the simple contribution  $ \ki W_{\vec{k}\vec{k\pr}}\delta (\DT \aen) 
 \DT \vec{f}  
$, which for $W$ constant can be put on the RTA form. Therefore, the RTA can qualitatively reproduce the correct result. Neither can there be any angularly constant terms in $f^{(E)}$ for $|N|\neq 1$. Here the RTA works well, in contrast to the $|N|=1$ case.


\begin{thebibliography} {99}

{\footnotesize

\bibitem{Kat} Y. K. Kato, R. C. Myers, A. C. Gossard, and D. D.  Awschalom, Science {\bf 306}, 1910 (2004). 

\bibitem{Sih} V. Sih, R. C. Meyers, Y. R. Kato, W. H. Lau, A. C. Gossard, and D. D.  Awschalom, Nature (London), {\bf 1}, 31 (2005).

\bibitem{Wun} J. Wunderlich, B. Kaestner, J. Sinova, and T Jungwirth, Phys. Rev. Lett. {\bf 94}, 047204 (2005).  

\bibitem{Zhao} H. Zhao, E. J. Loren, H. M. van Driel, and A. L. Smirl,  Phys. Rev. Lett. {\bf 96} 246601 (2006).

\bibitem{Val} S.O. Valenzuela and M. Tinkham,  Nature {\bf 442}, 176 (2006).

\bibitem{Kim}  T. Kimura, Y. Otani, T. Sato, S. Takahashi, and S. Maekawa, Phys. Rev. Lett. {\bf 98}, 156601 (2007).


\bibitem{Eng} H.-A. Engel, E. I. Rashba and B. I. Halperin, cond-mat/0603306v3.

\bibitem{Sch} J. Schliemann, Int. J. Mod. Phys. B {\bf  20}, 1015 (2006)

\bibitem{Sinitsyn} N. A. Sinitsyn, J. Phys. Cond. Mat. {\bf 20}  023201 (2008).

 
\bibitem{KarLut} R. Karplus and J. M. Luttinger, Phys. Rev. {\bf 95}, 1154 (1954).



\bibitem{DyaPer} M. I. Dyakonov and M. I. Perel, Sov. Phys. JETP Lett. {\bf 13} 467 (1971). 

\bibitem{Hir}  J. E. Hirsch,  Phys. Rev. Lett. {\bf 83}, 1834 (1999).


\bibitem{Zhang} S. C. Zhang, Phys. Rev. Lett. {\bf 85}, 393 (2000). 

\bibitem{Mur} S. Murakami, N. Nagaosa, and S. C. Zhang, Science {\bf 301}, 1348 (2003). 

\bibitem{SinPRL92} J. Sinova, D. Culcer, Q. Niu, N. A. Sinitsyn, T. Jungwirth, and A. H. MacDonald, Phys. Rev. Lett. {\bf 92}, 126603 (2004).





\bibitem{Mis} E. G. Mishchenko, A. V. Shytov, and B. I. Halperin, Phys. Rev. Lett. {\bf 93}, 226602 (2004). 

\bibitem{Kha} A. Khaetskii, cond-mat/0408136. 


\bibitem{Sug} N. Sugimoto, S. Onoda, S. Murakami and  N. Nagaosa, Phys. Rev. B {\bf 73}, 113305 (2006). 


\bibitem{Shy} A. V. Shytov, E. G. Mishchenko, H.-A. Engel and B. I. Halperin, 
Phys. Rev. B {\bf 73}, 075316 (2006). 

\bibitem{RaiGor} R. Raimondi, C. Gorini, P. Schwab, and M. Dzierzawa
Phys. Rev. B 74, 035340 (2006).


\bibitem{CulWin} D. Culcer and R. Winkler, PRB {\bf 76}, 245322 (2007). 











\bibitem{Ino2003} J. Inoue, G. E. W. Bauer, and L. W. Molenkamp, Phys. Rev. B {\bf 67}, 033104 (2003).

\bibitem{Ino2004}  J. Inoue, G. E. W. Bauer, and L. W. Molenkamp, Phys. Rev. B 70, 041303 (2004).


\bibitem{NomJun} K. Nomura, J. Sinova, T. Jungwirth, Q. Niu, and A.H. MacDonald, Phys. Rev. B {\bf 71 }, 041304(R) (2005). 

\bibitem{NomSin} K. Nomura, J. Sinova, N. A. Sinitsyn, and A. H. MacDonald, Phys. Rev. B {\bf 72},165316 (2005). 

\bibitem{RaiSch} R. Raimondi and P. Schwab, Phys. Rev. B {\bf 71},
033311 (2005). 

\bibitem{Dim} O. Dimitrova, Phys. Rev B {\bf 71}, 245327 (2005).
\bibitem{ChaLos} O. Chaleev and D. Loss, Phys. Rev. B {\bf 71}, 245318 (2005) 


\bibitem{Koe} M. Koenig, H. Buhmann, L. W. Molenkamp, T. L. Hughes, C.-X. Liu, X.-L. Qi, S.-C. Zhang, arXiv:0801.0901.   





\bibitem{Zub} D. Zubarev, V. Morozov and G. R\"opke, {\it Statistical Mechanics of Nonequilibrium Processes}, Vol 1\&2, Akademie Verlag (1996).

\bibitem{Ram} J. Rammer, {\it Quantum Transport Theory}, Perseus Books (1998).

\bibitem{Bro} P. Brouwer, {\it Theory of many-particle systems}, lecture notes, http://people.ccmr.cornell.edu/~brouwer/p654/ 

\bibitem{RamSmi} J. Rammer and H. Smith, Rev. Mod. Phys., {\bf 58}, 323 (1986).

\bibitem{ShiZha} J. Shi, P. Zhang, D. Xiao, and Q. Niu, Phys. Rev. Lett. {\bf 96}, 076604 (2006).

\bibitem{LipSpiVel} P. Lipavsky, V. Spicka and B. Velicky, Phys. Rev. B {\bf 34}, 6933 (1986).



%
\bibitem{CulWin2} D. Culcer and R. Winkler, arxiv:0807.3051. 


\bibitem{AusKat} M. Auslender and M. I. Katsnelson,
Phys. Rev. B {\bf 76}, 235425 (2007).

\bibitem{LipMorSpi} P. Lipavsky, K. Morawetz and V. Spicka, {\it Kinetic equations for strongly interacting dense Fermi systems}, Annales de Physique {\bf 26}, No. 01 (2001).

\bibitem{LagWil} D. C. Langreth and J. W. Wilkens, Phys. Rev. B {\bf 6}, 3189 (1972).


\bibitem{FujJalTan}  T. Fujita, M. B. A. Jalil and  S. G. Tan, arXiv:0903.2702. 


\bibitem{Hol} T. Holstein,   Ann. Phys. (Paris) {\bf 29}, 410 (1964).


\bibitem{Kov} A. A. Kovalev, K. Vyborny, J. Sinova, Phys. Rev. B 78, 041305(R) (2008).
 
\bibitem{DyaPer2} M. I. Dyakonov and V. I. Perel, Sov. Phys.-Solide State {\bf 13}, 3023 (1972). 

\bibitem{AveGol} N. S. Averkiev and L. E. Golub, Phys. Rev. B {\bf 60}, 15582 (1999).
\bibitem{GlaIvc} M. M. Glazov and E. L. Ivchenko, J. Supercond. {\bf 16}, 735 (2003).



}
\end{thebibliography}
\end{document}